\UseRawInputEncoding
\documentclass[10pt,letterpaper]{article}
\usepackage[top=0.85in,left=2.75in,footskip=0.75in,marginparwidth=2in]{geometry}
\usepackage[utf8]{inputenc}
\usepackage{cite}
\usepackage{amsmath,amssymb,amsfonts}
\usepackage{algorithmic}
\usepackage{graphicx}
\usepackage{textcomp}
\usepackage[hidelinks]{hyperref}
\usepackage{orcidlink}
\usepackage{subcaption}
\usepackage{array}

\usepackage{microtype}
\DisableLigatures[f]{encoding = *, family = * }

\raggedright
\usepackage{changepage}

\usepackage[aboveskip=1pt,labelfont=bf,labelsep=period,singlelinecheck=off]{caption}

\makeatletter
\renewcommand{\@biblabel}[1]{\quad#1.}
\makeatother

\usepackage{lastpage,fancyhdr,graphicx}
\usepackage{epstopdf}
\fancyheadoffset[L]{2.25in}
\fancyfootoffset[L]{2.25in}

\usepackage{color}

\definecolor{Gray}{gray}{.25}

\usepackage{graphicx}

\usepackage{sidecap}
\usepackage{wrapfig}
\usepackage[pscoord]{eso-pic}
\usepackage[fulladjust]{marginnote}
\reversemarginpar

\begin{document}
\vspace*{0.35in}

\begin{flushleft}
{\Large
\textbf\newline{A leadless power transfer and wireless telemetry solutions for an endovascular electrocorticography}
}
\newline
\\
{Zhangyu Xu \orcidlink{0000-0002-1536-2499}}\textsuperscript{1},
{Majid Khazaee \orcidlink{0000-0002-8257-2699} }\textsuperscript{2},
 {Nhan Duy Truong \orcidlink{0000-0003-4350-8026}}\textsuperscript{1,3},
Deniel Havenga\textsuperscript{1},
Armin Nikpour\textsuperscript{4},
{Arman Ahnood \orcidlink{0000-0002-0253-7579}}\textsuperscript{5},
{Omid Kavehei \orcidlink{0000-0002-2753-5553}}\textsuperscript{1,3,*}
\\
\bigskip
\bf{1} {School of Biomedical Engineering at The University of Sydney, Australia.}
\\
\bf{2} {Department of AAU Energy, Aalborg University, Denmark.}
\\
\bf{3} {BrainConnect Pty Ltd, Darlington, NSW 2008, Australia.}
\\
\bf{4} {Royal Prince Alfred Hospital, Camperdown NSW 2050, Australia.}
\\
\bf{5} {School of Engineering RMIT University, Melbourne, VIC 3000, Australia.}
\\
\thanks{The authors acknowledge the financial support from the Australian Research Council under Project DP230100019. The authors also acknowledge the support provided by the Sim4Life software support team.}
\bigskip
* Email: {omid.kavehei}@sydney.edu.au

\end{flushleft}

\section*{Abstract}
Endovascular brain-computer interfaces (eBCIs) offer a minimally invasive way to connect the brain to external devices, merging neuroscience, engineering, and medical technology. Achieving wireless data and power transmission is crucial for the clinical viability of these implantable devices. Typically, solutions for endovascular electrocorticography  (ECoG) include a sensing stent with multiple electrodes (e.g. in the superior sagittal sinus) in the brain, a subcutaneous chest implant for wireless energy harvesting and data telemetry, and a long (tens of centimetres) cable with a set of wires in between. This long cable presents risks and limitations, especially for younger patients or those with fragile vasculature. This work introduces a wireless and leadless telemetry and power transfer solution for endovascular ECoG. The proposed solution includes an optical telemetry module and a focused ultrasound (FUS) power transfer system. The proposed system can be miniaturised to fit in an endovascular stent.  Our solution uses optical telemetry for high-speed data transmission (over 2 Mbit/s, capable of transmitting 41 ECoG channels at a 2 kHz sampling rate and 24-bit resolution) and the proposed power transferring scheme provides up to 10mW power budget into the site of the endovascular implants under the safety limit. Tests on bovine tissues confirmed the system's effectiveness, suggesting that future custom circuit designs could further enhance eBCI applications by removing wires and auxiliary implants, minimising complications.

\section{Introduction}
\label{sec:introduction}
 Endovascular brain-computer interface (eBCI) systems have opened up new frontiers in human-machine interaction and have the potential to revolutionize the way we understand and treat neurological disorders~\cite{RN623,RN625,RN647,RN624}. It offers several benefits for high-fidelity chronic recordings of cortical neural activities. Firstly, it provides a minimally invasive approach, as it can be implanted through a blood vessel, reducing the risk of infection and tissue damage compared to traditional invasive methods. Additionally, the stent-electrode array allows for stable, long-term recordings due to its integration with the vessel wall, which results in fewer movement-related artifacts and improved signal quality~\cite{RN652,RN653}. This enables researchers and clinicians to gain deeper insights into brain function and develop more effective therapies for neurological disorders.
However, a significant limitation of the current stent-electrode array is the requirement of a long cable within the vessel to carry the signal to a recording device fixated in the chest. This arrangement presents a risk to the recipient, possibly leading to complications such as thrombosis, vessel injury, or infection~\cite{RN654,RN655}. Furthermore, the cable can pose a significant challenge in pediatric applications, as children constantly grow and develop~\cite{RN651,RN648,RN649}. The fixed cable length may not accommodate changes in the distance from the vessel to the recording device as the child grows, potentially leading to complications and the need for repeated surgical adjustments~\cite{RN656}.
To address these issues, it is crucial to develop a wireless telemetry module that satisfies the volume and dimension limits of the stent and can power itself by harvesting energy from an outside source. This wireless solution would improve eBCIs' safety level and expand the potential applications of the stent-electrode array across different age groups and clinical scenarios. However, several challenges have prevented the invention of the telemetry module, including:
\noindent\textbf{Area and volume limitation:} The telemetry module needs to be small enough to fit within the confines of the blood vessel without causing discomfort or impeding blood flow~\cite{RN623}. Designing a compact module that incorporates transducers for receiving energy and converting it into electrical power, as well as integrating components for modulation and encoding to transmit data, presents a significant challenge. This is further compounded by the need to integrate control and power management circuits, all within the limited available space.
\noindent\textbf{Power consumption limitation:} The module must be highly energy-efficient, as excessive power consumption will exceed the power delivery and harvesting budget and could generate heat, potentially damaging surrounding tissues~\cite{RN667}. Furthermore, for chronic recording, the power needs to be supplied and harvested continuously, which imposes constraints on the long-term safety concern and available energy format.
\noindent\textbf{High data rate requirement:} Capturing high-frequency neural activity and accommodating the data generated by a large number of electrodes necessitate a high data rate. A high sampling rate is required to accurately capture the high-frequency features of neural activity, while sufficient resolution of the analog-to-digital converter (ADC) is crucial to cover the dynamic range of the neural signals~\cite{RN658}. As the number of electrodes increases, the data rate multiplies accordingly, leading to a substantial amount of information that needs to be transmitted continuously. 
\noindent\textbf{Signal tissue penetration depth and absorption rate limitation:} Wireless transmission of neural data through biological tissues can be challenging due to signal attenuation and absorption~\cite{RN659, RN660}. The telemetry module must be able to transmit signals through various tissue layers while maintaining adequate signal strength and minimizing interference or distortion.
The development of a miniaturized wireless power and data module for eBCIs presents a multitude of challenges that require innovative engineering solutions. Moreover, the module must be biocompatible, robust, and reliable, considering it will operate in a highly sensitive and dynamic environment within the human body~\cite{RN663}. Addressing these challenges necessitates interdisciplinary collaboration, combining expertise in microelectronics, materials science, biomedical engineering, and signal processing. Ultimately, the successful development of a miniaturized wireless telemetry module for eBCIs holds the potential to revolutionize the field of implantable medical devices, offering new treatment options and improving the quality of life for patients with various neurological conditions.
This work introduces a novel solution that utilizes optical telemetry for data transmission and piezoelectric telemetry for energy harvesting. We address the size (cross-section area and volume), data rate, and power delivery and harvesting challenges by combining two types of technologies and taking advantage of each.
\section{Background}
\label{sec:Background}
Optical telemetry, particularly in implantable medical devices for neural recording, presents a compelling advancement in ensuring efficient data transmission while adhering to the stringent size and power constraints inherent to such applications~\cite{RN664}. The core principle of optical data telemetry hinges on transmitting data through optical signals, utilizing light, typically in the infrared or near-infrared spectrum, to encode and transmit information from the implantable device to an external receiver. This technology capitalizes on the intrinsic advantages of light as a medium, enabling high data rates, minimized power consumption, and a substantial reduction in the size of the telemetry module, thereby aligning well with the imperatives of modern implantable neural recording devices.
The optical data solution represents a powerful combination of three essential qualities: compact size, low power consumption, and high data rate. Together, these attributes effectively tackle the ongoing challenge of achieving a balance between size, power efficiency, and data transmission speed in implantable wireless devices~\cite{RN665}. The compact size of the optical telemetry module is a crucial innovation that allows for its integration into tiny implantable devices, thereby enhancing the capabilities of neural interfacing and recording technologies. Additionally, the module's low power requirements significantly reduce the challenges associated with power harvesting circuits and the limitations of the implant's power source.
Furthermore, the high data transmission speed provided by optical telemetry plays a crucial role in enabling the real-time transfer of neural data. This capability is essential for applications such as brain-computer interfaces, real-time monitoring of neurological disorders, and closed-loop neurostimulation systems~\cite{RN668}. This high-speed data transmission feature allows for the reliable and almost instant transfer of extensive neural information, facilitating prompt interventions and precise evaluations in both clinical and research environments. The wireless aspect of optical telemetry increases the flexibility and comfort of implantable neural recording devices, removing the requirement for external wires that might increase infection risks and limit patient movement. With these standout characteristics, optical data telemetry technology considerably advances the development of implantable medical devices, creating an environment that supports more advanced and patient-focused neural recording and interfacing solutions.
Piezoelectric energy harvesting is a burgeoning field within medical device technology, stemming from the unique ability of piezoelectric materials to convert mechanical energy into electrical energy when subjected to stress or strain~\cite{RN669}. This principle underpins the development of self-sustaining power systems within medical devices, capitalizing on the abundance of ambient mechanical energy sources such as body movements, blood flow, or external vibrations. A notable feature of piezoelectric energy harvesting technology is its compact form and high energy density, particularly advantageous for small-sized implantable or wearable medical devices~\cite{RN670}. These attributes alleviate the spatial constraints and power limitations traditionally associated with integrating batteries or other external power sources, thus significantly enhancing the feasibility and functionality of miniaturized medical devices.
Piezoelectric materials' compactness and high energy density are pivotal in advancing the size, power budget, and reliability of various medical devices. For instance, self-powered pacemakers~\cite{RN671}, insulin pumps~\cite{RN672}, and cochlear implants~\cite{RN673} have been developed by harnessing the piezoelectric phenomena to convert biomechanical energy from heartbeats, muscle movements, or external vibrations into electrical energy that powers these devices or deliver the desired function directly. Moreover, piezoelectric energy harvesting technology facilitates the continuous operation of critical monitoring systems and sensors, integral in chronic disease management and remote patient monitoring, by providing a steady and sustainable power supply~\cite{RN674}.
However, despite these advantages, there are several drawbacks associated with piezoelectric energy harvesting technology. One of the primary challenges is the relatively low energy conversion efficiency compared to other energy harvesting technologies, which may necessitate the incorporation of additional power management systems to ensure a consistent energy supply.  Furthermore, the biocompatibility and packaging method of piezoelectric materials within the human body remain areas of active investigation, as piezoelectric material will raise safety concerns when implanted in the human body, and packaging could potentially compromise power harvesting efficiency~\cite{RN675, RN676}. Despite these challenges, piezoelectric materials' compact nature and high energy density leave them as one of the best candidates for powering eBCI by harvesting energy within the stent.
Focused ultrasound (FUS) waves offer new opportunities for piezoelectric energy harvesting for implantable device. This technique uses ultrasonic waves, concentrated at a specific point inside the body, to transmit energy through the skin and tissue directly to the device. Unlike general ultrasound, which disperses energy over a wider area, FUS concentrates energy precisely at a specific target area. This targeted approach allows for more efficient energy transfer, minimizing the dissipation of energy through non-targeted tissues and reducing the potential for unintended heating or damage. Still, the related ultrasound-piezoelectric issues need further investigations, such as pressure amplitude at deep tissue~\cite{RN677}, toxic material's connection with tissue~\cite{RN678}\cite{RN679}, and relatively low power output~\cite{RN680}. 
\section{Method}
\label{sec:Method}
\begin{figure}[ht]
    \centering
    \includegraphics[width= 0.95\textwidth]{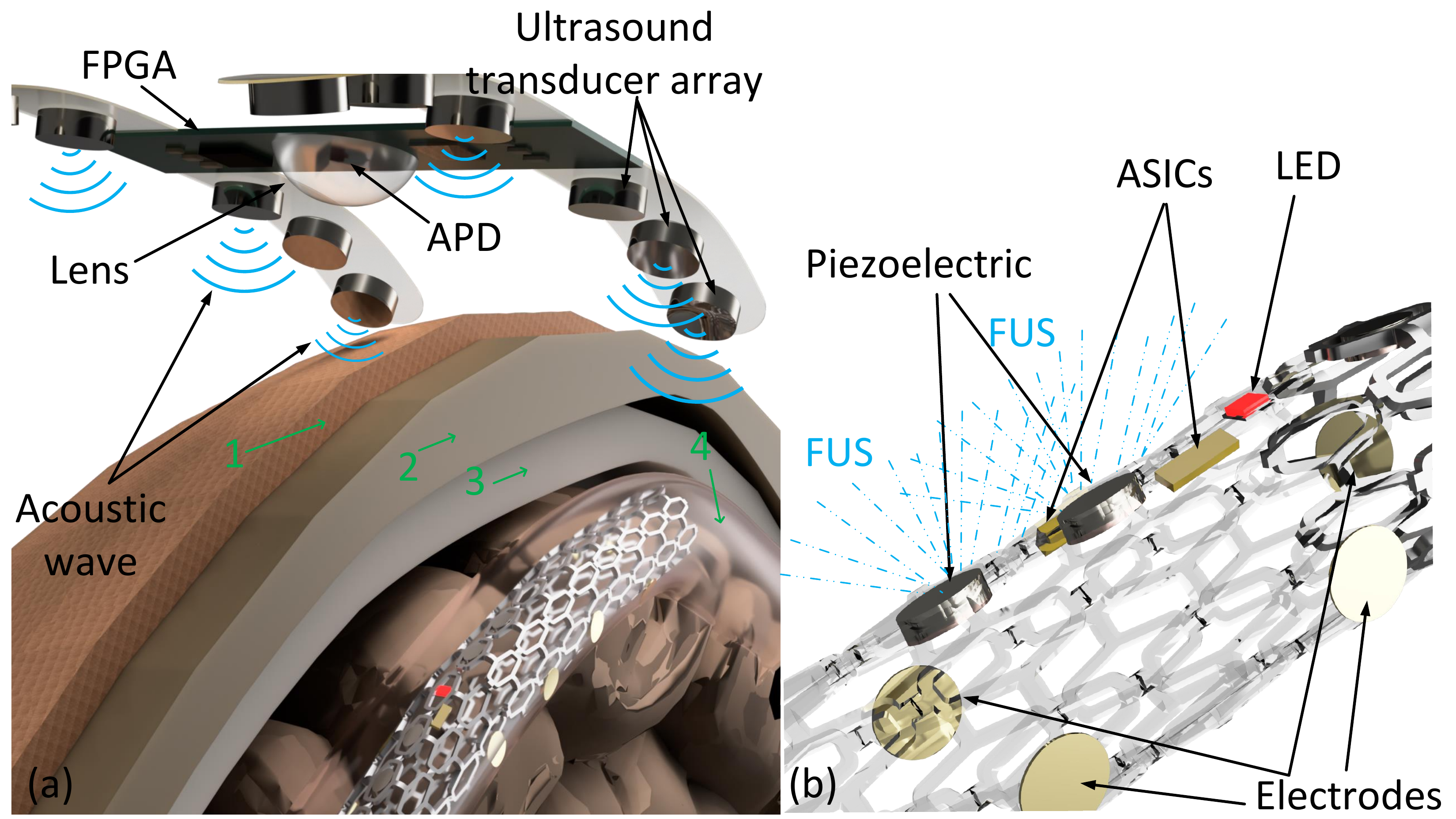}
    \caption{(a) Illustration of the system. In sub-figure a, the green arrow shows the tissues. (1) Skin tissue, (2) Bone tissue, (3) Dura mater, (4) Superior sagittal sinus. The system will have two parts. An implantable stent and an external device. All electrical components will sit in the stent in the superior sagittal sinus for the implantable part. The external device sits over the scalp and aligns with the implant. In the external device, an avalanche photodiode is used to collect optical signals from the implant, and an FPGA is used to decode the optical data. The ultrasound transducer array in the external device will generate focused ultrasound that delivers energy to the stent. Sub-figure b shows the stent with functional components. The sensing electrodes are on the stent for sensing electrical signals from the cortex. Three piezoelectrics sit in the stent to convert energy from focused ultrasound (FUS) to power the circuit. The optical transmitter, an 810 nm wavelength LED (shown in red), sits in a space within the stent. Two ASICs with control and sensing circuits, energy harvesting and power management circuit and LED driver sit in other spaces within the stents. }
    \label{fig:pw_fg1}
\end{figure}
Here, we present a wireless solution for eBCIs that meet the requirements of low-volume, low-energy, high data rate and can be powered wirelessly. This solution includes an optical data telemetry module that uses a single light-emitting diode (LED) as the transducer. With a simplified circuit design, the data telemetry module can transmit data at a high bit rate while consuming low power and occupying a minimal area. Our experiments showed that we can achieve 5 Mbit/s telemetry data rate and consume less than 4 mW of power with 7 mm bone and 10 mm soft tissues (fat, muscle and skin) between the transmitter and the receiver. The result indicates that the proposed module can transmit neural recording from a 32-channel electrode array with each channel sampled at 9.7~kHz with 16-bit resolution. The proposed solution includes a piezoelectric energy harvesting unit with multiple small piezoelectric power harvesters attached to the stent. Our simulation result showed we could harvest a maximum of 3~mW from one piezoelectric before the delivered acoustic pressure reaches the safety limit.  With six or more piezoelectrics, we can easily have more than a 10~mW power budget to supply all the sensing, data telemetry, and potential stimulation circuits.
Figure~\ref{fig:pw_fg1} shows the conceptual design of the proposed module working with an eBCI. In the implant part, three small piezoelectric harvesters are embedded in the stent to power the circuits. An 810nm wavelength LED sits in the stent as the transducer for data telemetry. Two small ASICs embedded in the stent are the sensing, control, data communication, and power management circuits.
Benefiting from the small size of the piezoelectric components and the LED, the proposed module has a transducer volume of less than 2~mm$^3$ (One LED and One piezoelectric). The module has significant potential for miniaturization ability. The proof-of-concept testing board created in this work uses discrete components, and the total volume is less than 14~mm$^3$ (excluding PCB), indicating that with future ASIC development, the volume of the module can be smaller than 4.5~mm$^3$. The entire module can be easily embedded into a stent with this volume.
To validate the proposed design, we did analysis, simulation, and proof-of-concept experiments according to the following parameters of success:
\subsection{Optical channel analysis}
\label{sec:Optical channel analysis}
\begin{figure}[ht]
    \centering
    \includegraphics[width=0.9 \textwidth]{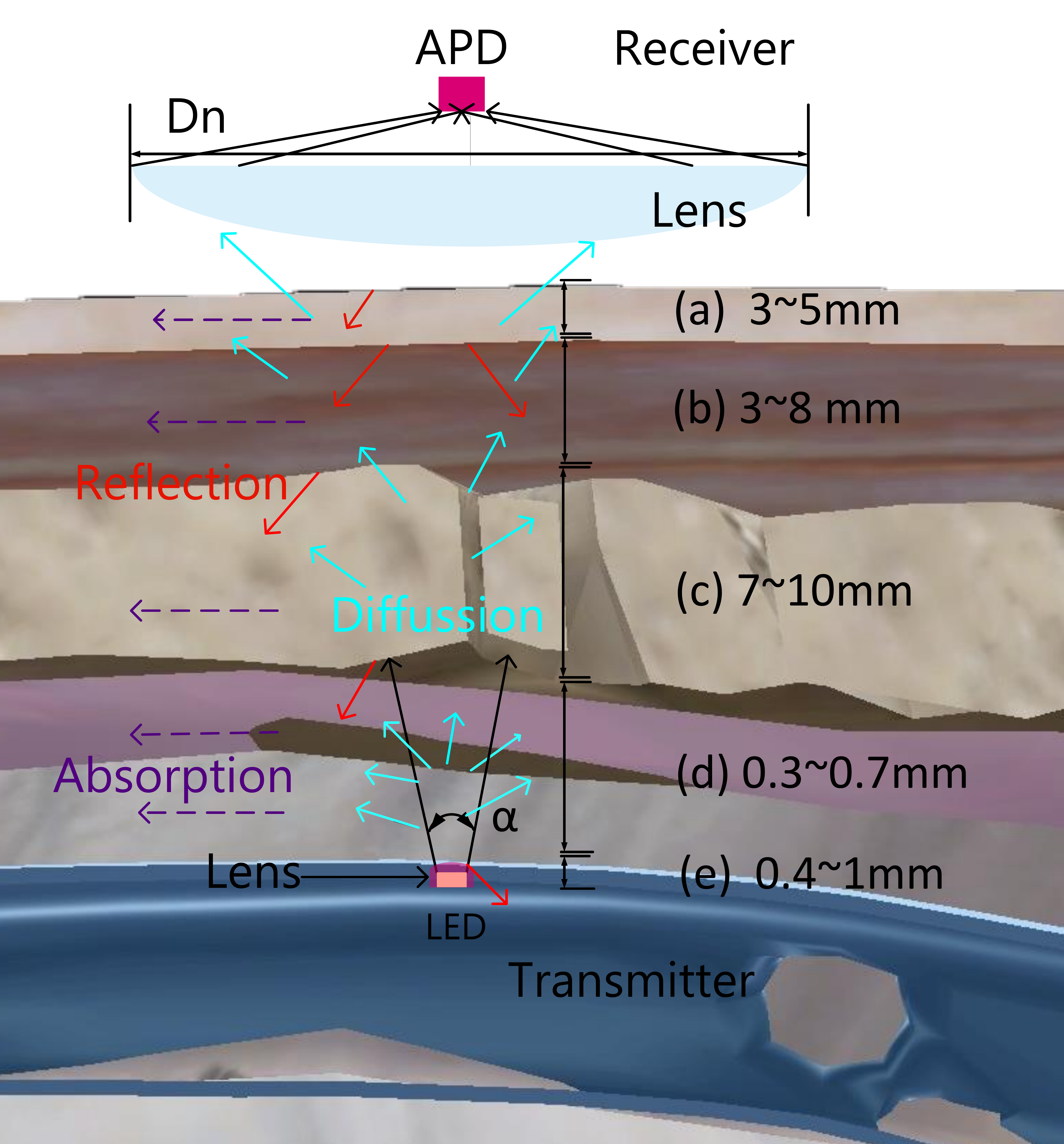}
    \caption{(a) Skin, (b) Connective tissue, Galea aponeurotica, Loose areolar connective tissue, Periosteum, (c) Skull, (d) Dura Mater, (e) Superior sagittal sinus Wall. Dn is the diameter of the lens. Alpha is the emitting angle of the LED, which the emitting lens can control.
    The arrows represent the diffusion (blue), reflection (red), and absorption effect (purple).}
    \label{fig:optical_ana}
\end{figure}
In the passage of 810~nm wavelength light through the human head's layers, distinct optical behaviors are observed (see Figure~\ref{fig:optical_ana}). The Superior sagittal sinus wall, characterized by a thin endothelial lining, shows low absorption and diffusion at this wavelength, with reflection influenced by refractive index differences between blood and adjacent tissues. The Dura Mater, a dense collagen and elastin membrane, exhibits moderate absorption and high diffusion, with reflection arising from refractive index disparities. The skull, comprising cortical and trabecular bone, has low absorption, moderate diffusion due to its porous structure, and reflection at tissue interfaces. Connective tissue, rich in collagen, elastin, and proteoglycans, demonstrates low absorption and moderate diffusion, with reflection at tissue boundaries. The skin, with its epidermal melanin and dermal blood vessels, shows moderate absorption and high diffusion, with reflection at the interfaces of its layers. Overall, the optical properties at 810nm involve varying absorption, diffusion, and reflection, primarily influenced by blood, melanin, and water absorption, the fibrous nature of tissues, and structural complexities facilitating light scattering and reflection at tissue interfaces.
\subsection{System design}
\label{sec:System design}
\begin{figure}[ht]
    \centering
    \includegraphics[width=0.9\textwidth]{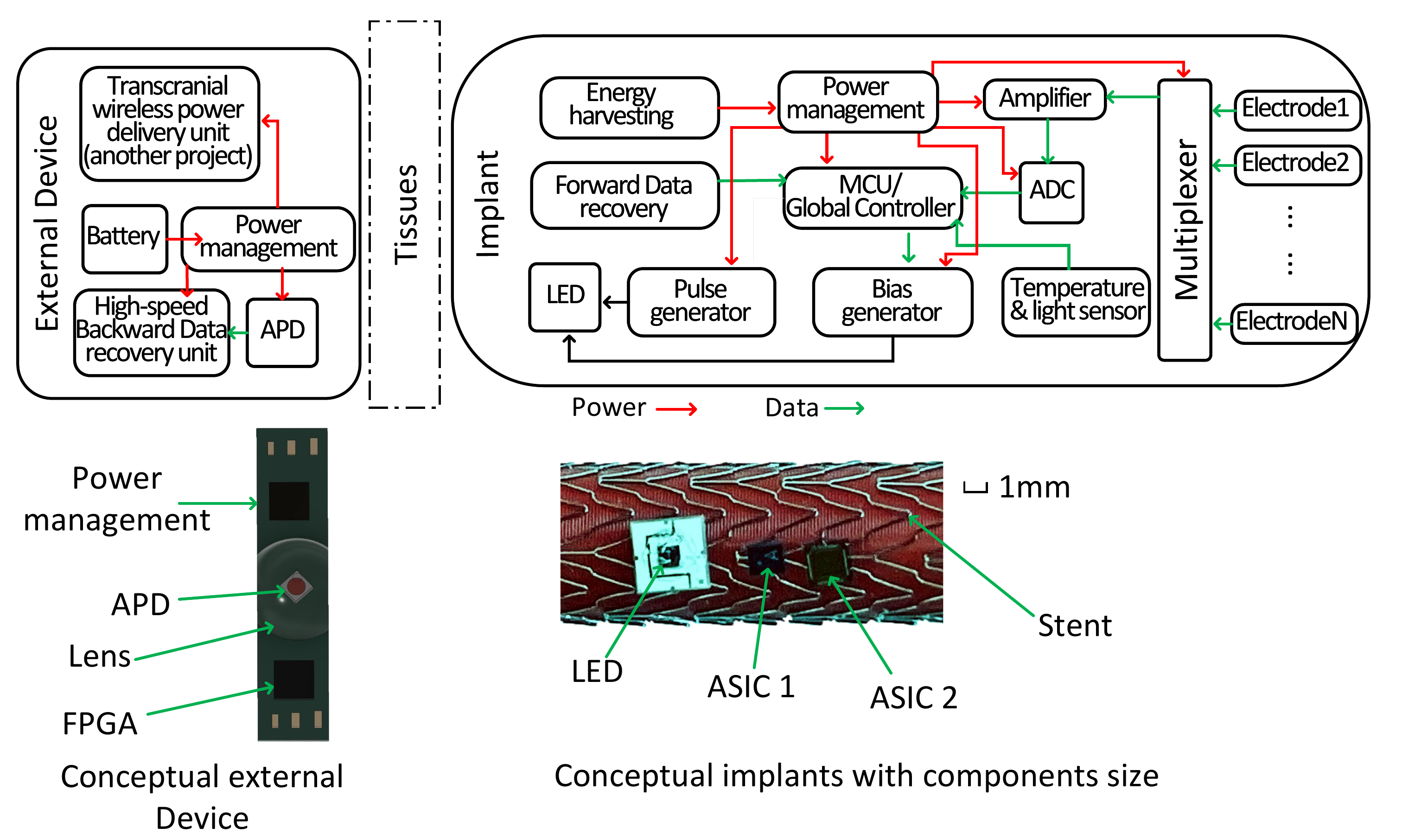}
    \caption{System block diagram. 
    The top left corner shows the blocks in the external device. The data recovery will be implemented on FPGA and the wireless power delivery unit will be described in another work. The upper right part shows the blocks in the implant. The lower right part shows the potential components' size with an actual stent. The LED is a commercially available one and has a big substrate. We need a customized smaller substrate to meet our purpose.}
    \label{fig:system_dia}
\end{figure}

Figure~\ref{fig:system_dia} shows the system's block diagram and the device's conceptual design. 
To eliminate the need for a long cable connecting the brain to the chest, our plan involves integrating energy harvesting, sensing, control, power management, and communication modules directly within the stent itself. An external device will be positioned over the implant site outside the body for seamless communication and power delivery.
To accommodate the stent's dimensions, we propose utilizing two application-specific integrated circuits (ASICs) to house the required circuitry. The first ASIC will encompass the global controller and sensing circuitry, including a multiplexer, amplifier, analog-to-digital converter (ADC), and temperature sensor. The second ASIC will contain the energy harvesting, power management, bias, and pulse generator circuits. By making the circuit into two smaller ASICs, the chips can be more easily adapted to fit the constrained dimensions of the stent and minimize the impair to the surrounding tissue or blood flow, which provides more flexibility in component placement and optimizes the use of the stent's internal space. Separating power-intensive components (such as energy harvesting and power management circuits) from sensitive analog components (like amplifiers and ADCs) can reduce the risk of noise interference and improve signal integrity. This can result in higher accuracy and better overall performance of the implant.
The external device comprises three primary units: a data receiver unit, a wireless power delivery unit, and a power and battery management unit. Given its exceptional sensitivity and high-speed performance, we use an avalanche photodiode (APD) as the optical data receiver. An FPGA (field-programmable gate array) is used in our bench experiment for data recovery due to its flexibility and processing capabilities, which far exceed our needs. The power management unit will be responsible for battery management, supplying the bias voltage required for the APD, and providing power to the FPGA.
\subsection{System powering plan}
\label{sec:System powering plan}
The proposed solution for powering the implant in the brain incorporates an external transducer array and an array of minute piezoelectric materials embedded within the stent itself. The transducer array, situated externally over the head, is meticulously engineered to generate focused ultrasound waves. These waves are specifically directed toward the stent’s location within the brain, ensuring precision in energy delivery. The transducer array comprises numerous individual transducers, each of which can be controlled independently. This allows for fine-tuning the ultrasound beam in terms of intensity, focus, and direction, ensuring that the energy is delivered accurately to the stent while minimizing exposure to the surrounding brain tissue.
\begin{figure}[ht]
    \centering
    \includegraphics[width=0.9\linewidth]{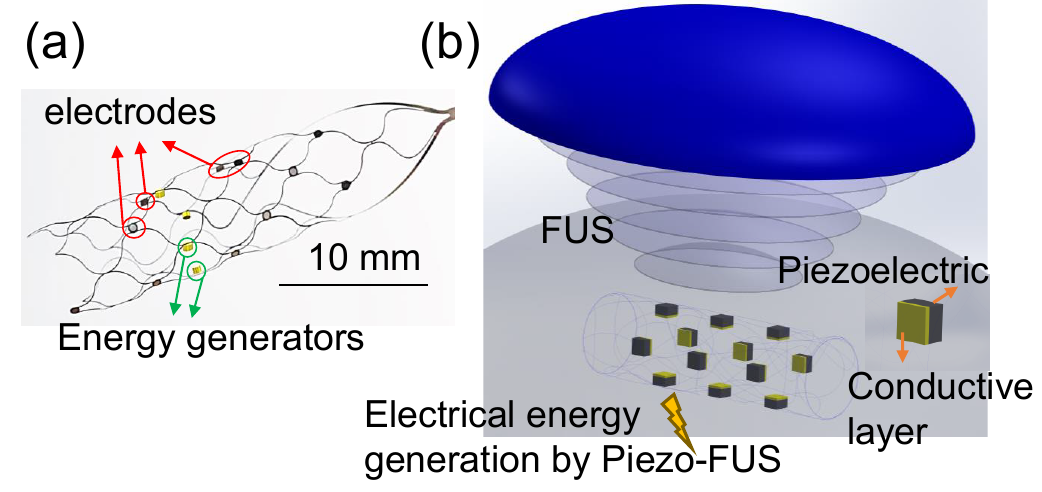}
    \caption{The structure of the piezoelectric harvester //Illustrating the process of electrical energy generation using Piezo-FUS: A focused ultrasound beam (blue ellipse) targets a stent embedded with a thin layer of piezoelectric material (highlighted in orange). Upon interaction with the ultrasound, the piezoelectric layer converts the acoustic energy into electrical energy, represented by the lightning bolt symbol.}
    \label{fig:PictureModelDescription}
\end{figure}
The stent, positioned within a blood vessel in the brain, is designed to be minimally invasive while providing the necessary support to the vessel. Embedded within the stent are small piezoelectric materials, which have the capability to convert the acoustic energy from the ultrasound waves into electrical energy. A lead-free potassium-sodium niobate with high piezoelectric coefficient $d_{33}$ is employed~\cite{RN681}. These materials are selected for their high energy conversion efficiency and biocompatibility, ensuring that they function effectively within the body without inducing adverse reactions. The piezoelectric materials are connected to the energy harvesting circuit, which converts the harvested electrical energy into a stable power supply for the other circuits.

The focused ultrasound (FUS) is generated by a group of ultrasound generators in the external unit. To ensure the safety of the ultrasound energy, we not only set intensity limits for each of the ultrasound generators but also carefully control the intensity of the FUS waves to prevent any potential damage to the surrounding brain tissue. Additionally, we plan to use the telemetry channel to include feedback mechanisms, which can let the external unit dynamically adjust the output power, ensuring that the energy delivery and conversion processes are operating efficiently and safely.

Overall, this solution provides a novel method of powering endovascular implants in the brain, utilizing the synergy between FUS and piezoelectric materials to create a self-sufficient system that enhances patient monitoring and intervention capabilities while adhering to stringent safety standards.
\begin{figure}[ht!]
    \centering
    \includegraphics[width=0.90\linewidth]{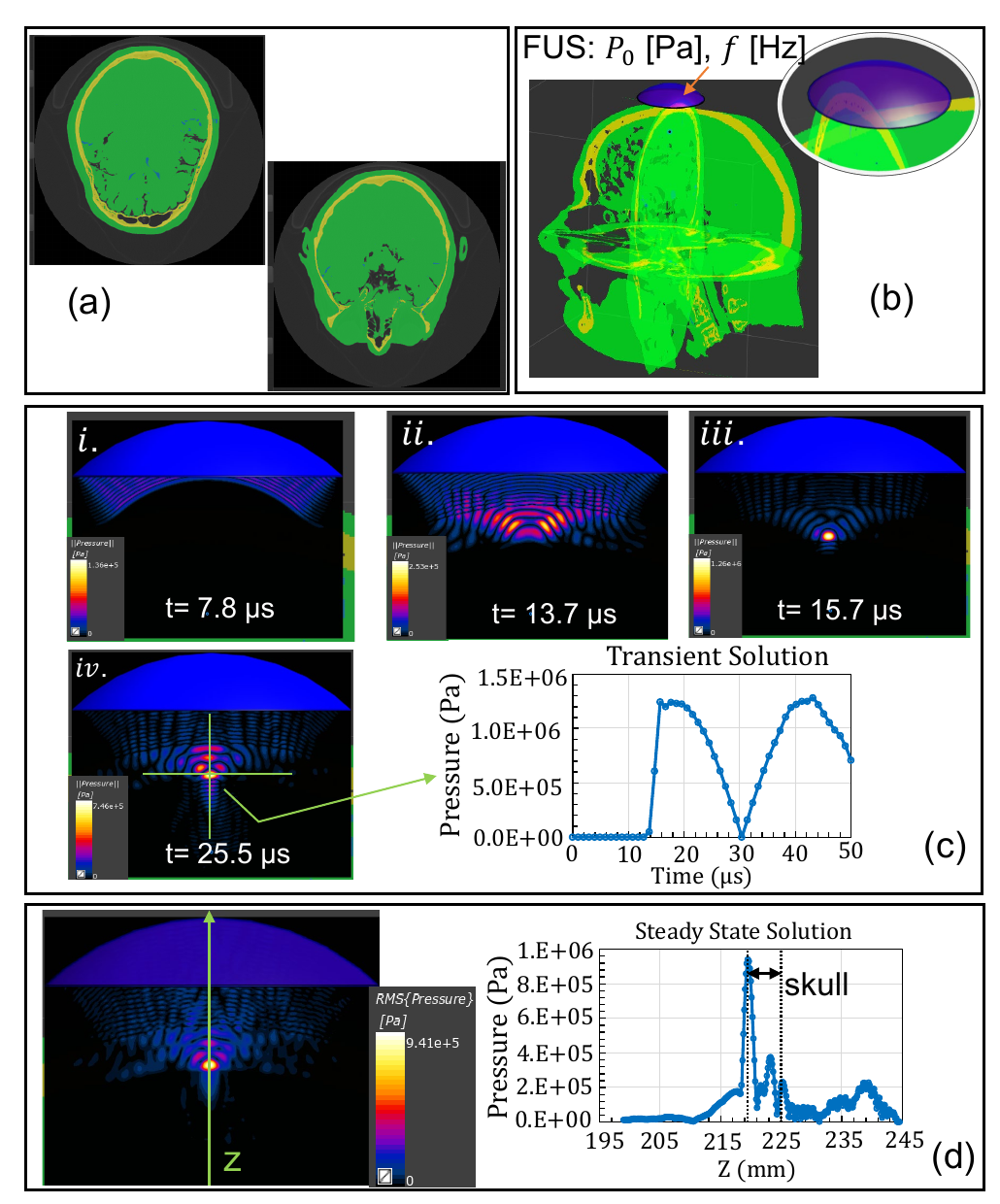}
    \caption{Modelled focused ultrasound/brain, (a) real CT (computed tomography) scan images, (b) modelled human head and a single spherical focused ultrasound transducer, (c) $P_0$ acoustic wave propagation through the head by the transient solution and the transient acoustic pressure at the scalp, and (d) the steady state solution of acoustic pressure and the acoustic pressure at the focal line.}
    \label{fig:PictureAcousticPressure}
\end{figure}
Figure~\ref{fig:PictureModelDescription} shows the structure of piezoelectric elements that we proposed in this work. The piezoelectric material surface is covered by a metallic conductive layer and an electrode connection with the stent. The existence of multiple piezoelectric elements ensures a constant and high-power generation.

Figure~\ref{fig:PictureAcousticPressure} visually represents the modeling and simulation process of focused ultrasound propagation through the human brain. Initially, in part (a), real CT (computed tomography) scan images from the Visible Human project by iSEG tool \cite{RN682},   provide a detailed cross-sectional view of the human head, showcasing various layers and anatomical details. This initial simulation is $P_0=30$ kPa and $f=$ 1 MHz ultrasound waves. Part (b) offers a modeled representation of the human head by Sim4Life \cite{RN650} alongside a single spherical focused ultrasound transducer, highlighting the positioning and biological interaction. This transducer is responsible for generating the focused ultrasound waves. Moving on to part (c), the figure delves into the dynamics of acoustic wave propagation through the head. It showcases the transient solution, illustrating how these waves change quickly. Additionally, it presents the transient acoustic pressure experienced at the scalp, which provides insight into the intensity of the waves as they first interact with the head. Lastly, in part (d), the figure shifts to the steady-state solution, capturing the point where the acoustic waves stabilize and don't change significantly over time. The acoustic pressure at the focal line is also displayed, indicating the targeted region's precise pressure within the brain where the ultrasound is most concentrated. This comprehensive figure, therefore, offers a holistic view of how focused ultrasound interacts with and propagates through the intricate structures of the human head and brain. The steady-state acoustic pressure along the Z-axis is shown in Figure~\ref{fig:PictureAcousticPressure}~(d), clearly showing the effect of skull bone on the acoustic pressure concentration; therefore, it is expected that the critical temperature rise occurs at the skull.
\begin{figure}[ht!]
    \centering
    \includegraphics[width=0.90\linewidth]{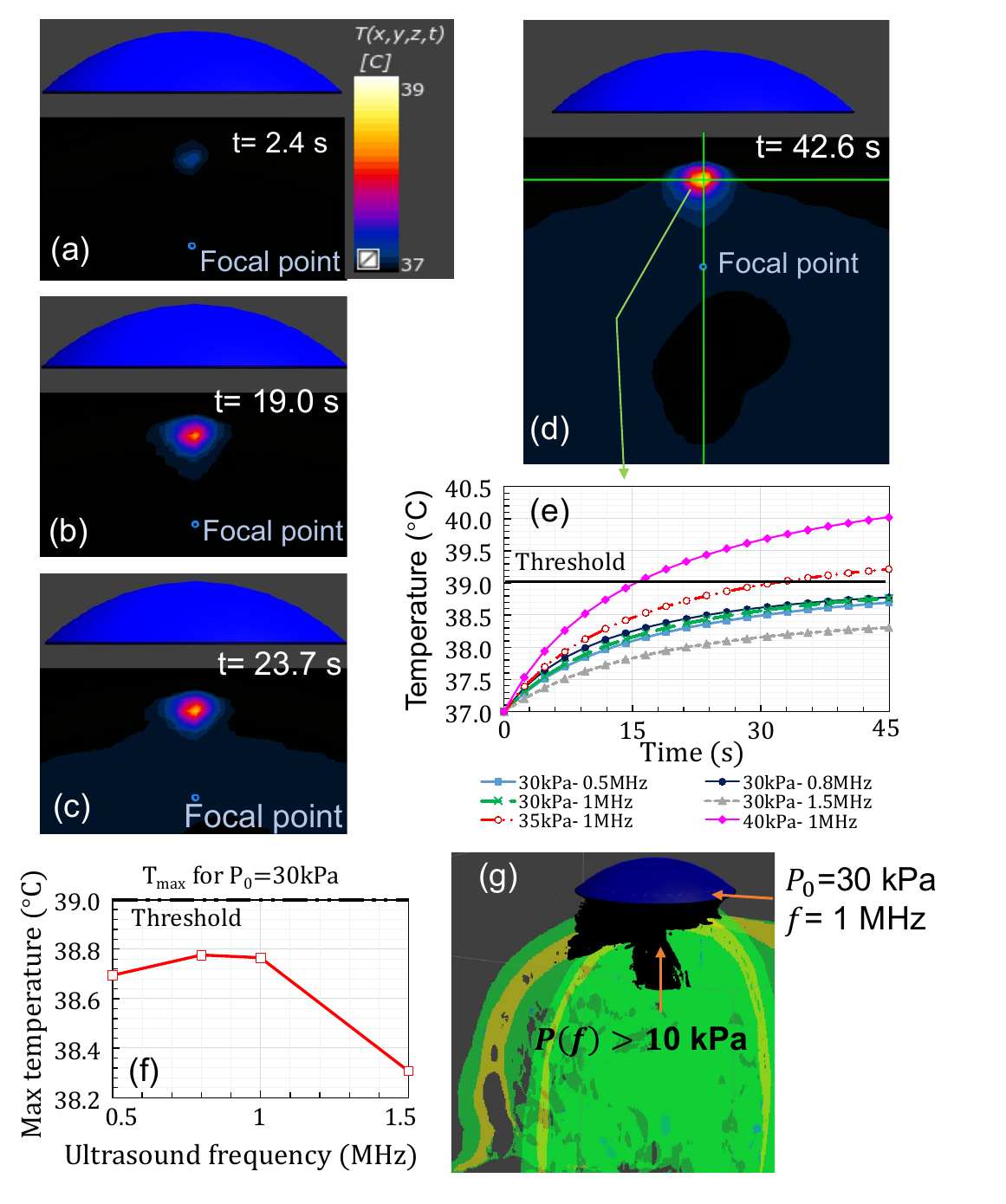}
    \caption{ Ultrasound-thermal analysis (a)-(d) The transient temperature contours over time with $P_0=$ 30 kPa and $f=$ 1 MHz, (e) temperature at the top of the skull over time for different acoustic settings, (f) maximum temperature versus ultrasound frequency, and (g) The steady-state iso-volume pressure $>$ 10 kPa inside the brain }
    \label{fig:PictureAcousticThermal}
\end{figure}
Acoustic-thermal analysis is carried out to meet the long-term thermal safety criteria. Transient ultrasound-thermal analysis for $P_0=$ 30 kPa and $f=$1 MHz case is shown in Figure~\ref{fig:PictureAcousticThermal} (a) to (d). Due to the acoustic pressure concentration on the skull, the skull temperature rises to the critical temperature. Multiple ultrasound parameters variations are set with the limitation of temperature rise below 2 ℃. The temperature rise for all frequencies is always below 2 ℃ for $P_0=$ 30 kPa; thus, the acoustic pressure is used for the safe ultrasound.
\subsection{Experiment setup}
\label{sec:Experiment setup}
\begin{figure}[ht]
     \centering
     \begin{subfigure}[t]{0.3\textwidth}
         \centering
         \includegraphics[width=\textwidth]{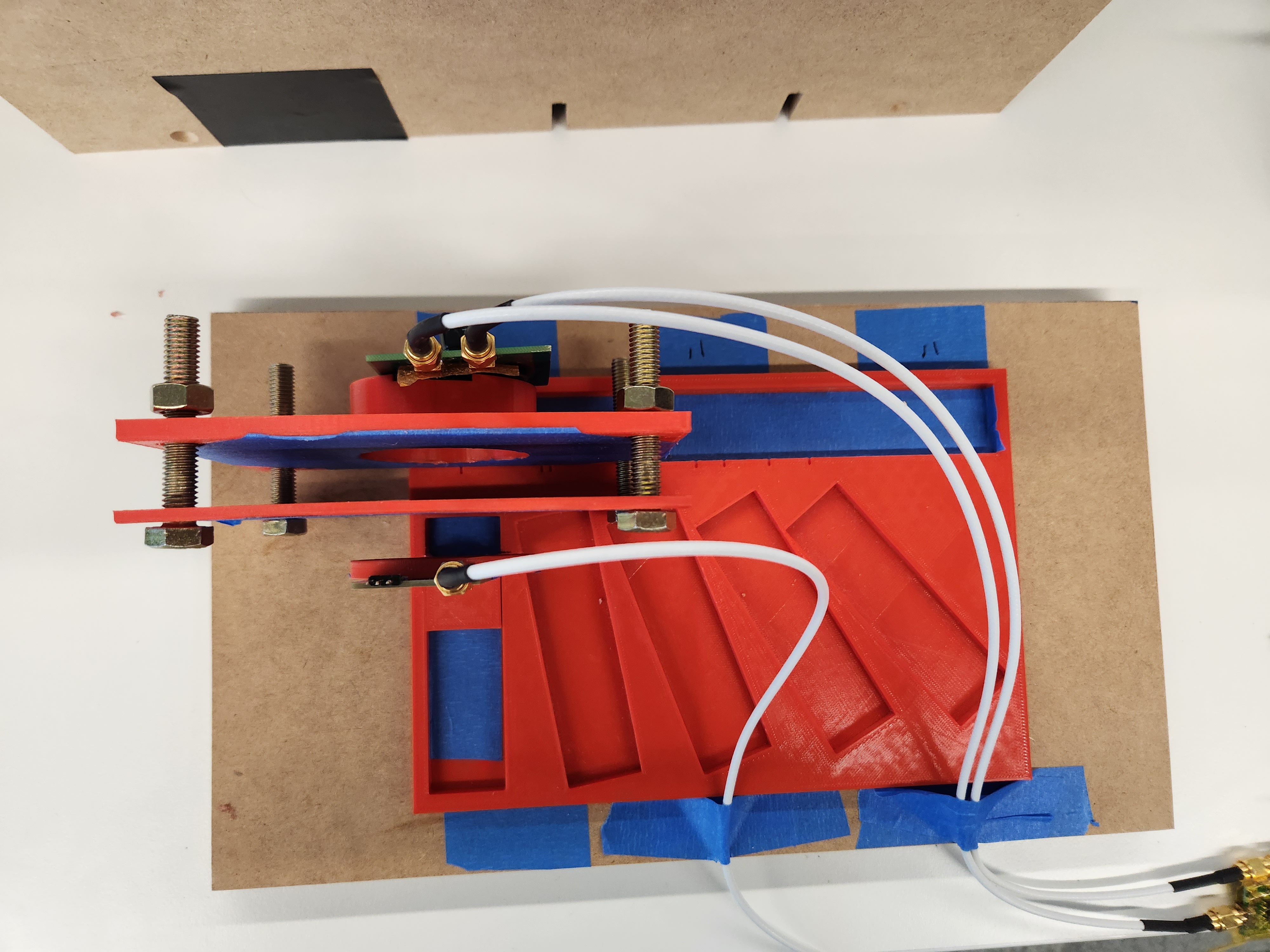}
         \caption{}
     \end{subfigure}
     \hfill
     \begin{subfigure}[t]{0.3\textwidth}
         \centering
         \includegraphics[width=\textwidth]{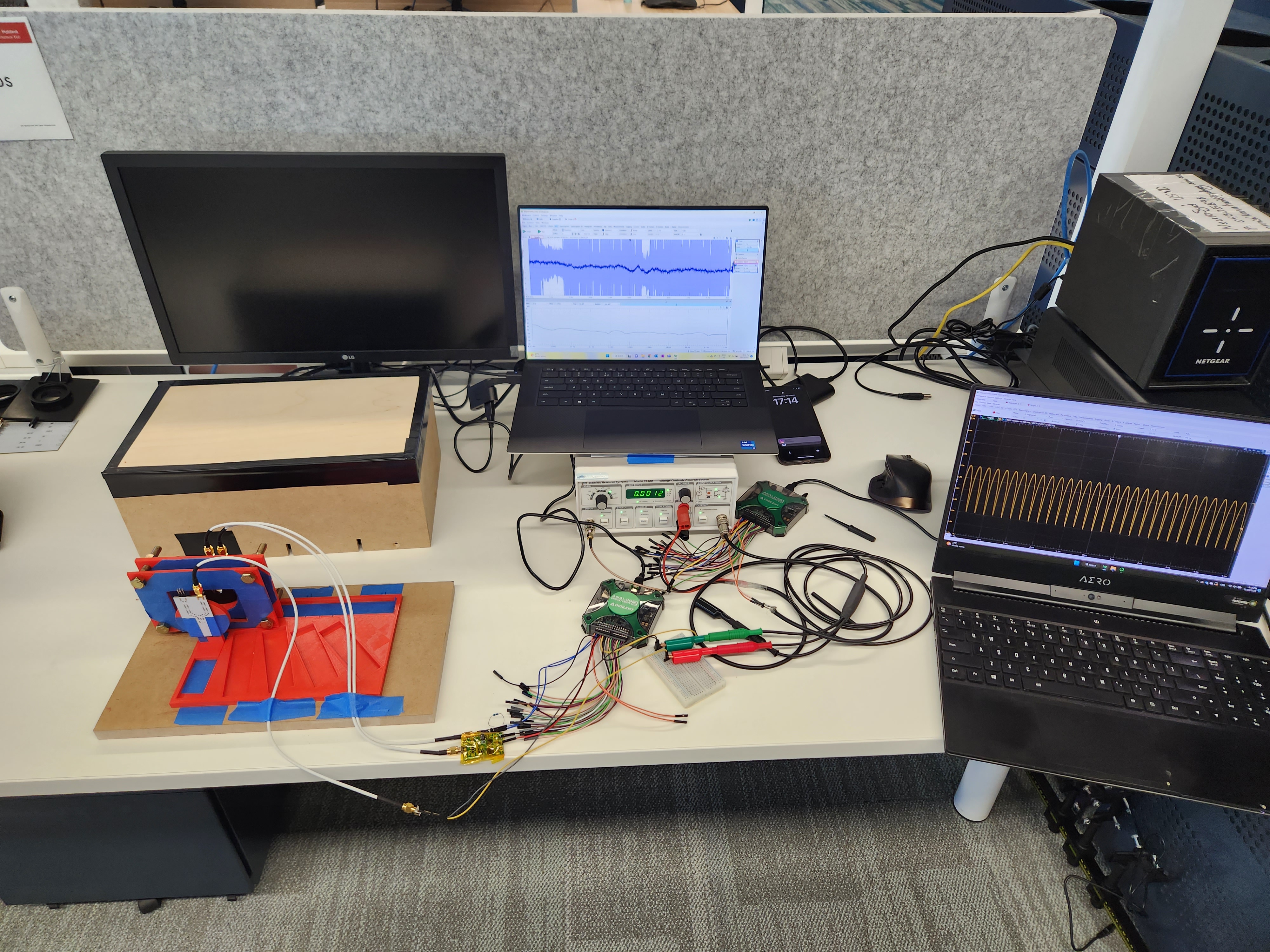 }
         \caption{}
     \end{subfigure}
     \hfill
     \begin{subfigure}[t]{0.3\textwidth}
         \centering
         \includegraphics[width=\textwidth]{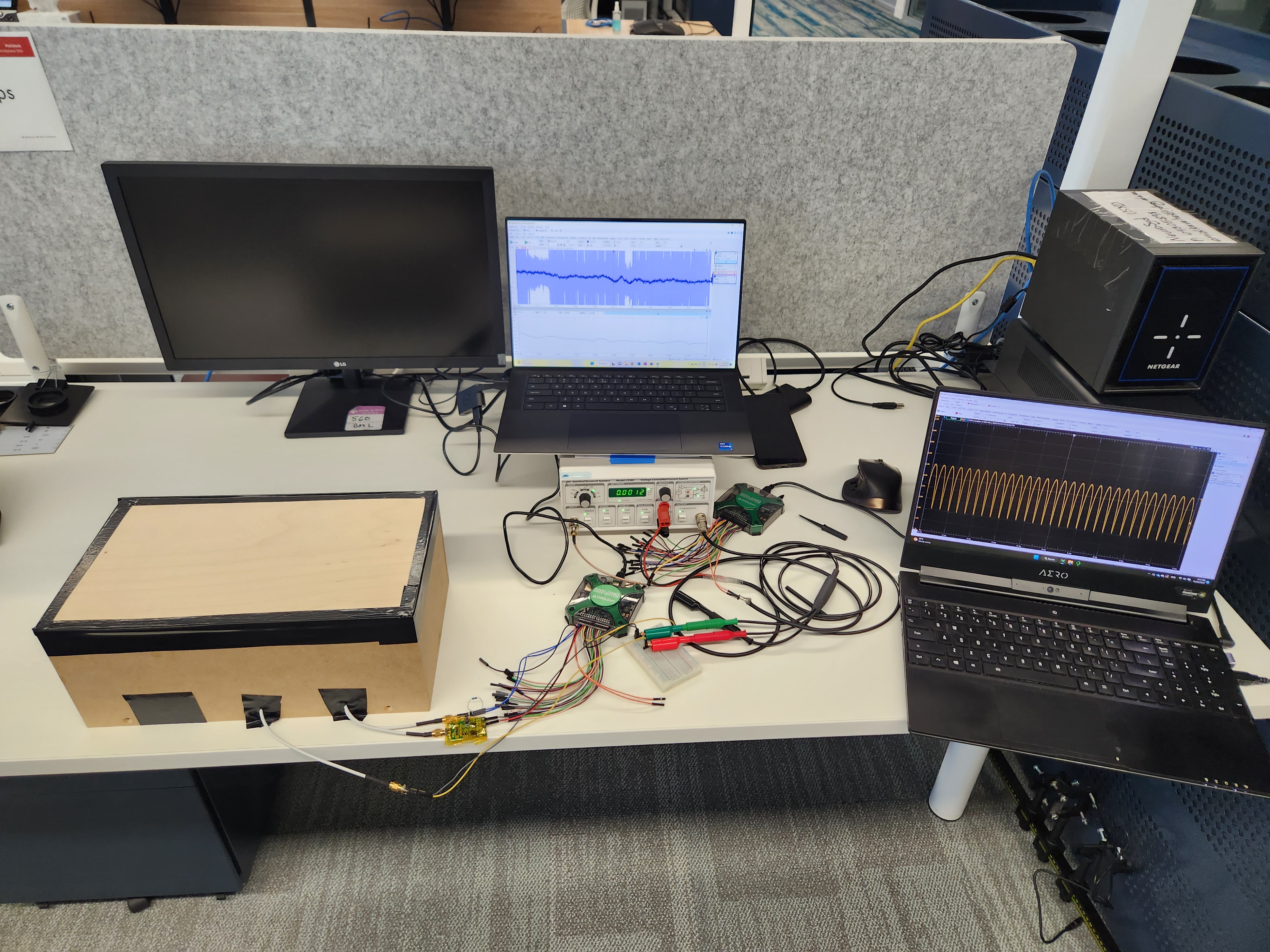}
         \caption{}
     \end{subfigure}
        \caption{This figure shows the set-up of the experiments (a) shows the 3D-printed stage and holder that held the biological specimens between an avalanche photo-diode and 810~mm LED which were aligned at a distance of 50~mm. (b) this shows the overall setup where the computer on the far left generates data for the transmitter, while the centre computer is part of the receiver that decodes the receiving data (c) is the fixture covered by a box to seal out any other environmental light while performing the experiment.}
        \label{fig:Experiement_1_Setup}
\end{figure}

In order to evaluate the design and performance of the telemetry module and its associated system, we manufactured custom-printed circuit boards (PCBs) featuring discrete components for experimental purposes. Using 3D printing technology, we created holders to securely position the PCBs and the biological tissue samples during testing. These 3D-printed holders facilitated optimal optical alignment and precise control over distance and potential misalignment settings. A visual representation of the experimental setup can be found in Figure~\ref{fig:Experiement_1_Setup}.

\begin{figure}[ht]
     \centering
     \begin{subfigure}[t]{0.3\textwidth}
         \centering
         \includegraphics[width=\textwidth]{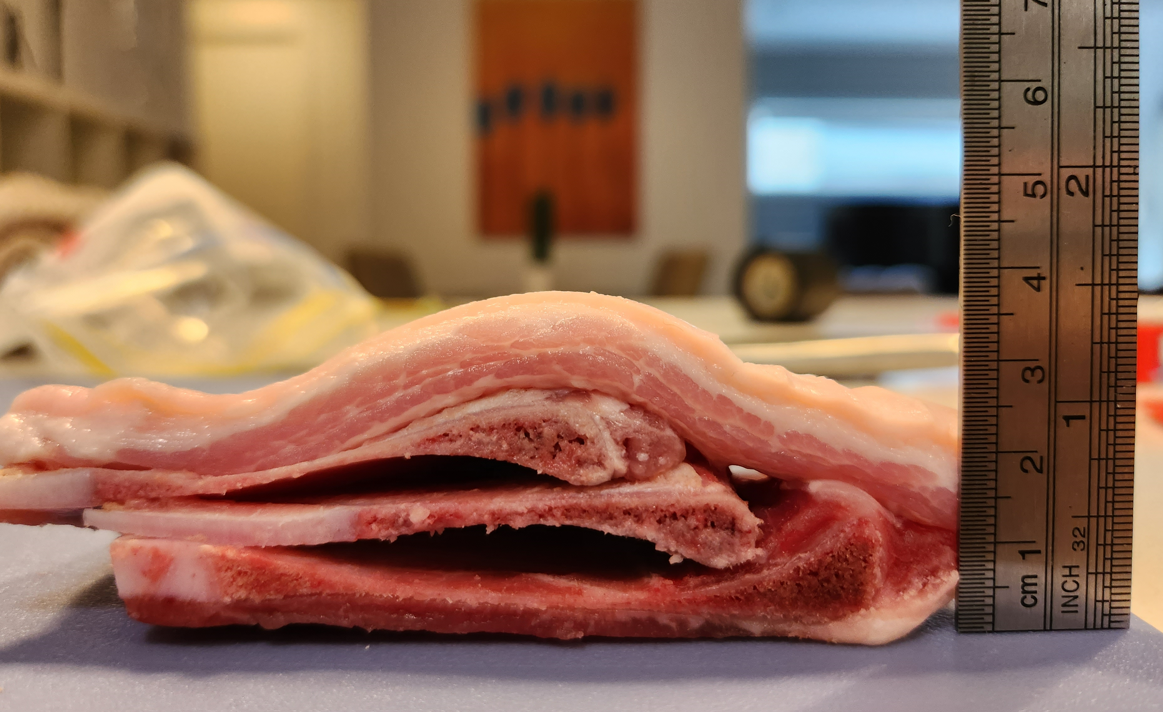}
         \caption{}
     \end{subfigure}
     \hfill
     \begin{subfigure}[t]{0.3\textwidth}
         \centering
         \includegraphics[width=\textwidth]{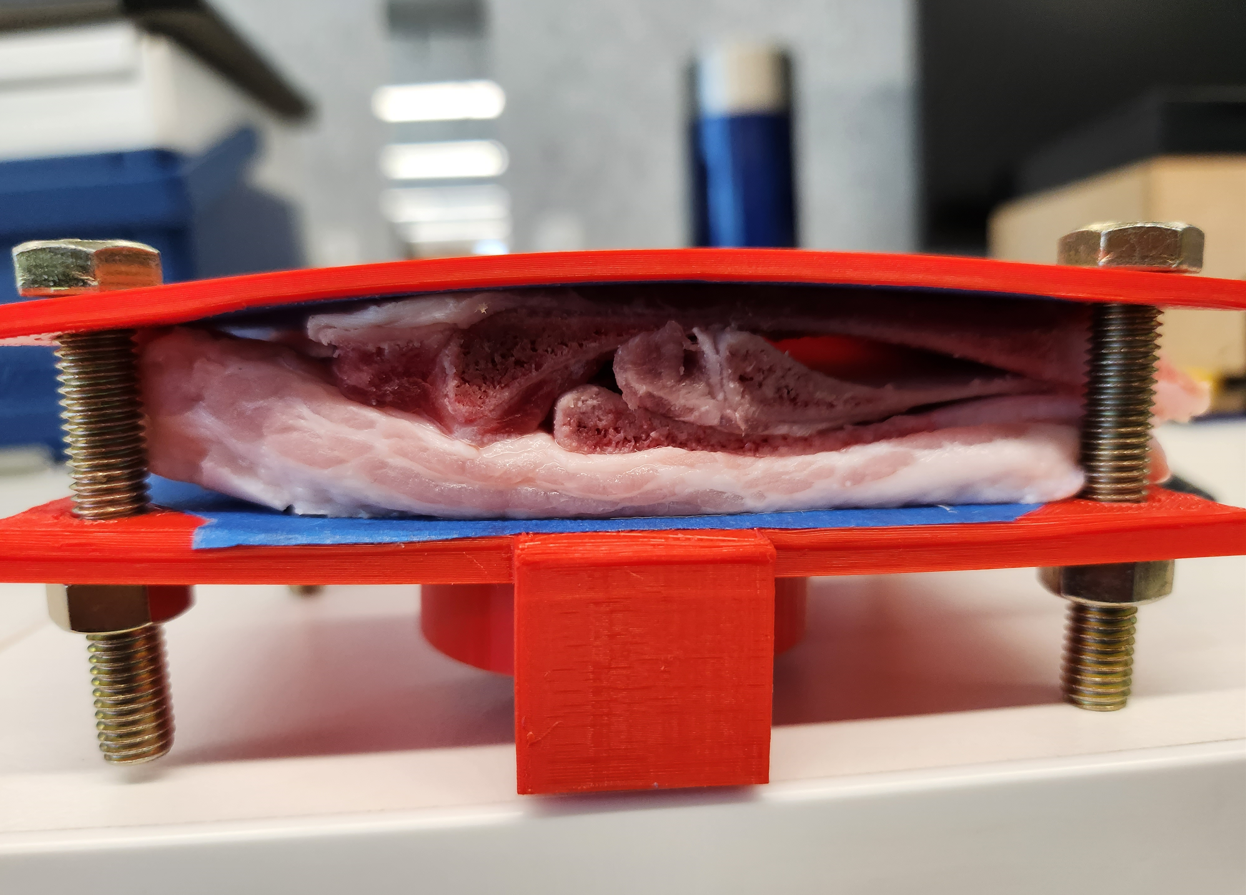}
         \caption{}
     \end{subfigure}
     \hfill
     \begin{subfigure}[t]{0.3\textwidth}
         \centering
         \includegraphics[width=\textwidth]{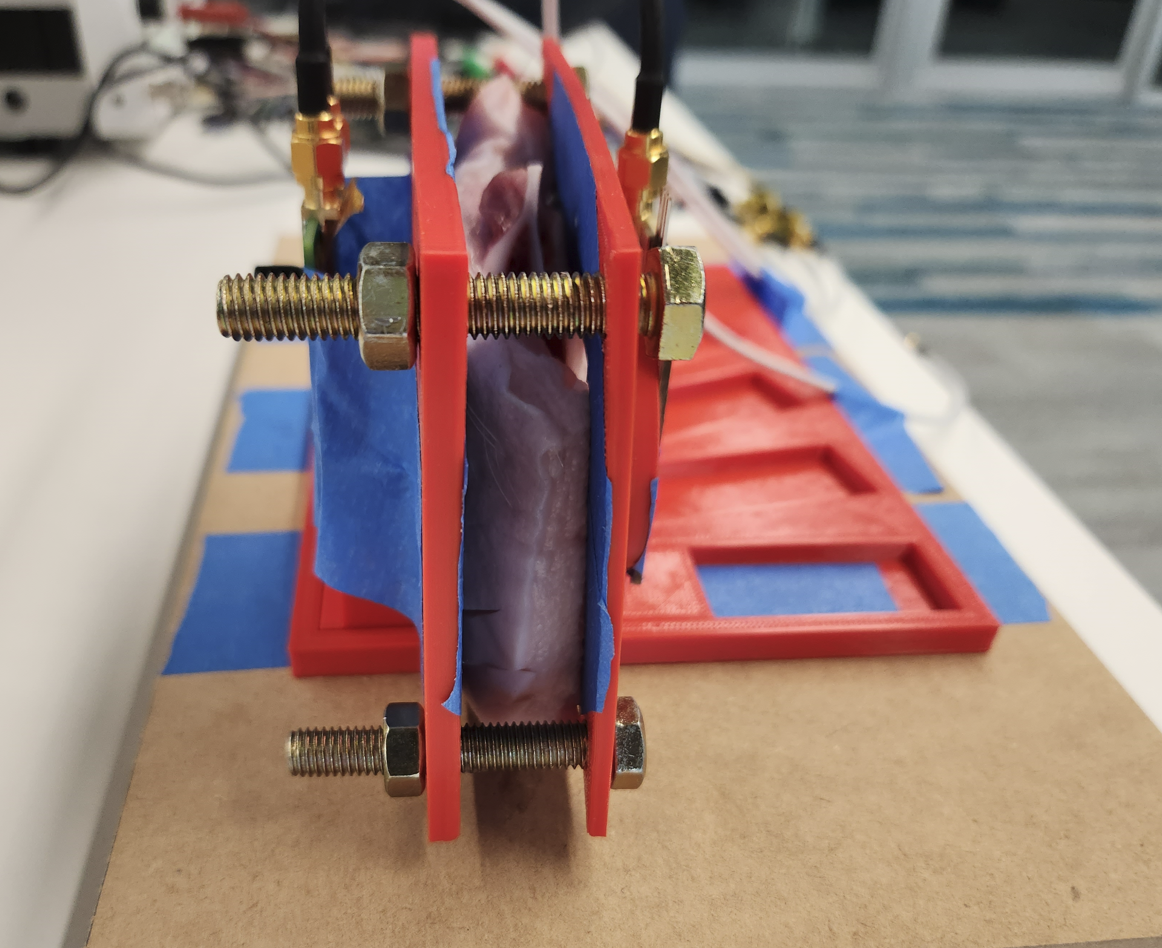}
         \caption{}
     \end{subfigure}
        \caption{This figure shows the tissue we used for the experiment. (a) The tissues and their thickness. The top layer is skin tissue with subcutaneous fat. The lower three layers are bones. During the experiments, we stack different layers of bones to get different thicknesses of bone. (b) We use a 3D-printed frame to hold the tissues and use screws to put pressure to make the tissue firmly connected. (c) The experiment setup with tissues in between. The middle is the tissues. The left side is the receiver PCB and the right side is the transmitter.   }
        \label{fig:Experiement_1_Sample_Prep_1}
\end{figure}

The tissue samples employed in the testing process were freshly obtained bovine skin, complete with subcutaneous tissue and bone. The 3D-printed frame was used to maintain a firm grip on the tissue samples throughout the experiment. A depiction of the prepared tissue samples is provided in Figure~\ref{fig:Experiement_1_Sample_Prep_1}.

\subsection{Testing protocol}
\label{sec:Testing protocol}
\begin{figure}[ht!]
    \centering
    \includegraphics[width=0.9\textwidth]{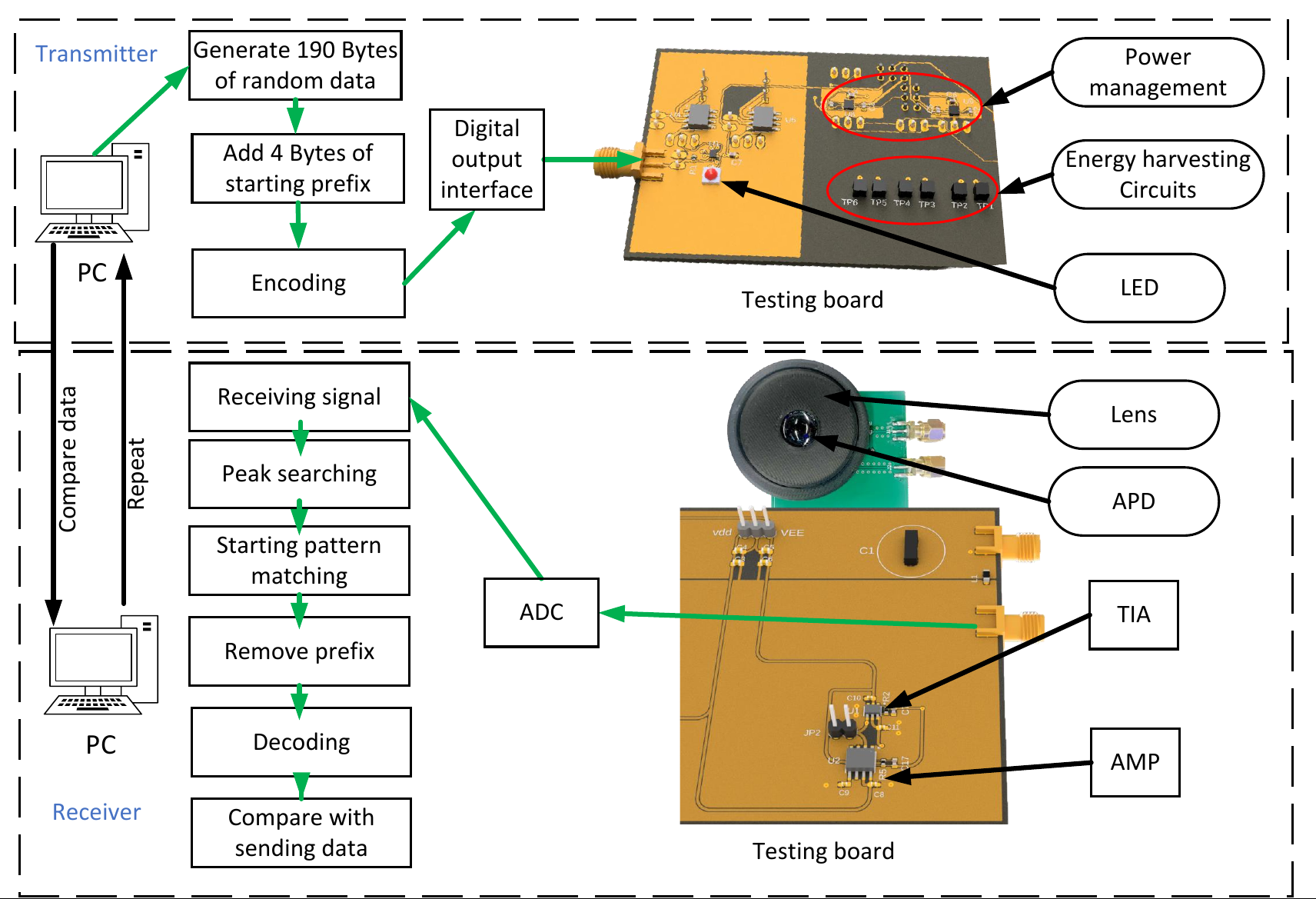}
    \caption{The upper portion of the diagram represents the transmitter. A computer running a program generates 190 bytes of random data at a time, limited by the buffer size of the digital I/O device. The program then adds a 4-byte prefix consisting of repeating fixed data to indicate the beginning of the transmission. The data is subsequently encoded into a pulse signal using either pulse width or pulse-density modulation. A digital I/O device generates the pulse and sends it to the testing board, which in turn produces optical pulses based on the input pulse signal. The lower portion of the diagram represents the receiver. Our testing board includes an analog front-end that converts the received optical pulses into electrical signals. An ADC digitizes the signal and sends it to a computer. Depending on the modulation scheme used, the program on the computer performs different operations. For pulse-density modulation, it conducts peak searching and marks the pulses on the timeline. For pulse width modulation, it uses a duty cycle searching and labeling algorithm to determine the pulse width. Once the signal is converted into binary format, the program searches for the repeated prefix pattern. When an entire segment of the prefix is matched, the program considers this as the starting point of the data and removes the prefix accordingly. After decoding, the program compares the recovered signal with the original transmitted data to verify the data transfer and calculate the data error rate. }
    \label{fig:testing_pro3}
\end{figure}
To confirm the effectiveness of data transmission via the prototype device, we established a comprehensive testing protocol outlined as follows:
Initially, we generate a sequence of random numbers using a random number generator. To this sequence, we add a 32-bit prefix as a starting indicator for data transmission. This prefix consists of four repeats of an 8-bit fixed pattern, which helps align the signal for decoding at the receiver's end.
Following this, the combination of random numbers and the prefix is encoded using either pulse-density modulation or pulse-width modulation, depending on the test iteration and the data transmission speed being tested. We use an Analog Discovery 2 (AD2) device and its Application Programming Interface (API) to create a signal with binary voltage levels, simulating the digital control and data transmission circuits of our proposed system. The output signal from the AD2 is then connected to a custom testing board designed to include our optical pulse generator circuits built from individual components. In response to the data sent, the board's Light Emitting Diode (LED) emits optical pulses, marking the completion of the transmission test sequence.
At the receiving end, an avalanche photodiode captures the optical pulses and converts them into electrical signals. These signals are then amplified by a trans-impedance amplifier, followed by a cascaded operational amplifier, and fed into another Analog Discovery 2 (AD2) unit.
In this second AD2, the Analog-to-Digital Converter (ADC) converts the analog signals into digital format, making them ready for transfer to a personal computer (PC). This PC uses an adaptive peak-searching algorithm to identify the pulses. After finding the peaks, the algorithm decodes the signal into a binary sequence based on set pulse-density thresholds.
The software begins by searching for the start indicator. Due to the lack of synchronization between sending and receiving, a single 8-bit pattern is used as the search criterion. Once a matching pattern is found, the data following the last bit of the prefix is extracted and shortened to 1520 bits.
The software then compares this 1520-bit sequence with the originally transmitted data to calculate the Bit Error Rate (BER). Subsequently, the receiving PC signals the transmitting PC to generate a new set of random numbers for the next test.
A comprehensive testing setup was deployed to evaluate the BER at specified transmission speeds accurately. This setup can create random images, encode them using the described method, and then decode them to conduct a thorough bit-level comparison between the sent and received image data, counting the total erroneous bits found. To establish a dependable BER metric, this testing protocol was performed continuously for 24 hours at each set transmission speed. The tests were carried out indoors with ambient lighting kept around 200 Lux, without direct sunlight exposure, and without using any environmental light shielding during the BER tests, to ensure the system's effectiveness under typical operating conditions.

Figure~\ref{fig:testing_pro3} shows the testing protocol and how we verify the data transmission has been successful.
This testing protocol ensures a comprehensive evaluation of the prototype device's data transfer capabilities, considering synchronization, signal decoding, and error rate analysis.
The power consumption is measured by connecting a voltage source to the power management part directly and measuring the input current, so the power consumption already includes the power consumption of the power management unit, LED driver and the driver for data.

\begin{figure}[ht!]
    \centering
    \includegraphics[width=0.95\linewidth]{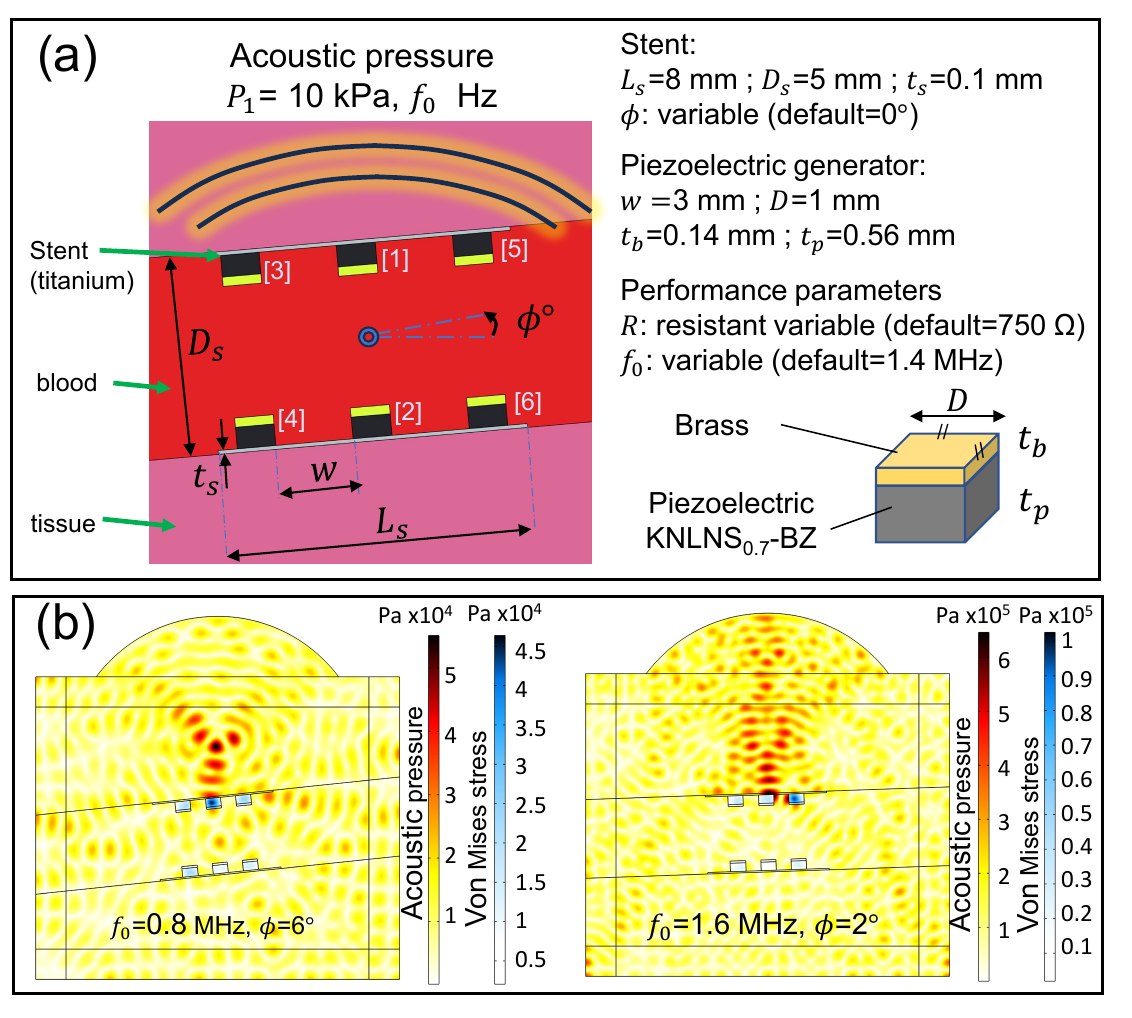}
    \caption{The acoustic-electro-mechanical simulation of energy harvesting system, (a) model parameters of piezoelectric generators, biomedical tissues, and acoustic pressure, (b) acoustic pressure in the tissue surrounding the piezoelectric generators and Von Mises stress on the piezoelectric generators for two settings.}
    \label{fig:PicturePressureWaves}
\end{figure}

To test the power harvesting plan, we combined COMSOL~\cite{RN626} and Sim4Life simulation software~\cite{RN650} to test the harvested energy under the safety limit. Figure~\ref{fig:PicturePressureWaves} (a) shows the model parameters and piezoelectric element characteristics. Figure~\ref{fig:PicturePressureWaves} (b) illustrates two primary results. The ultrasound frequency plays a vital role in piezoelectric energy generation as it affects the mechanical stress in the piezoelectric. The stent orientation is crucial as it affects the ultrasound-solid interactions. 
\section{Result}
\label{sec:Result}
We did data transfer experiments using the protocol described in Figure~\ref{fig:testing_pro3}.  The result is shown in Table~\ref{table:dresult}.

\begin{table}[ht!]
\centering
\begin{tabular}{
  |>{\centering\arraybackslash}m{1.6cm}
  |>{\centering\arraybackslash}m{2.2cm}
  |>{\centering\arraybackslash}m{1.4cm}
  |>{\centering\arraybackslash}m{1.8cm}
  |>{\centering\arraybackslash}m{1.8cm}|
}
\hline
\textbf{Data rate (Mbit/s)} & \textbf{Tissue} &  \textbf{Modul-ation} &  \textbf{Power consumption (mW)} &  \textbf{Efficiency (nJ/bit)} \\
\hline
0.5 & Bone (5mm) $+$ Skin (7mm) & PWM & 1.1 & 2.3 \\
\hline
1 & Bone (5mm) $+$ Skin (7mm) & PWM & 1.3 & 1.3 \\
\hline
2 & Bone (5mm) $+$ Skin (7mm) & PWM & 1.8 & 0.9 \\
\hline
5 & Bone (5mm) $+$ Skin (7mm) & PWM & 2.7 & 0.54 \\
\hline
1 & Bone (5mm) $+$ Skin (7mm)& PDM & 1.4 & 1.4 \\
\hline
3 & Bone (5mm) $+$ Skin (7mm)& PDM & 2.7 & 0.9 \\
\hline
2 & Bone (8mm) $+$ Skin (7mm) & PWM & 2.1 & 1.05 \\
\hline
5 & Bone (8mm) $+$ Skin (7mm) & PWM & 3.4 & 0.68 \\
\hline
3 & Bone (8mm) $+$ Skin (7mm) & PDM & 2.4 & 0.8 \\
\hline
2 & Bone (10mm) $+$ Skin (7mm)&  PWM & 2.6 & 1.3 \\
\hline
5 & Bone (10mm) $+$ Skin (7mm)&  PWM & 3.8 & 0.76 \\
\hline
3 & Bone (10mm) $+$ Skin (7mm) & PDM & 2.9 & 0.96 \\
\hline
\end{tabular}
\caption{Testing results for pulse width modulation (PWM) and pulse-density modulation (PDM).}
\label{table:dresult}
\end{table}

From the result, we can find that the data transfer efficiency is higher when using pulse width modulation. Still, PWM requires more precision timing to ensure the optical pulses can be decoded at the receiver side. When we use pulse-density modulation, we can relax more on timing and simplify the driver circuit for data.
\begin{figure}[ht!]
    \centering
    \includegraphics[width=0.6\linewidth]{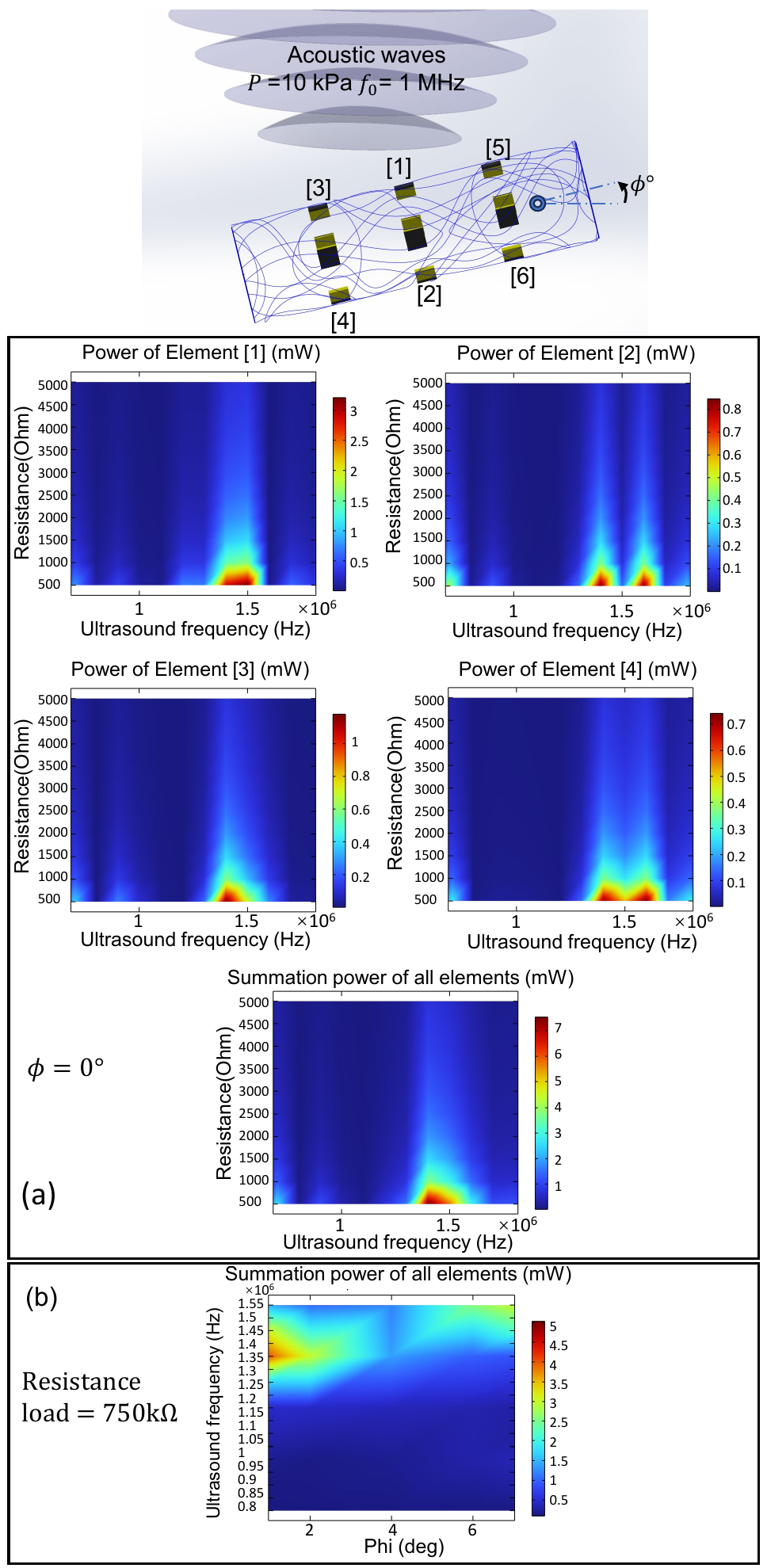}
    \caption{Simulation of piezoelectric power generation with FUS acoustic pressure, (a) piezoelectric power output from element number [1] to [4], and power summation for all elements. Power comparison versus ultrasound frequency and resistance load connection, and (b) the effect of stent-system orientation on the power output.}
    \label{fig:PressurePower}
\end{figure}

Figure~\ref{fig:PressurePower} shows the power harvesting result of the proposed power harvesting plan. We can see that a single piezoelectric generates the maximum power of 3.0 mW, and with the six elements, the sum of power output is 10.0 mW. With rotated piezoelectric elements, a high power output can be generated by tuning the frequency of FUS waves. These results illustrate the tunding of the frequency and placement of the FUS transducer for each piezoelectric; we can get the sum of the maximum single piezoelectric and reach a high level of power generation.

\begin{figure}[ht!]
    \centering
    \includegraphics[width=0.9\textwidth]{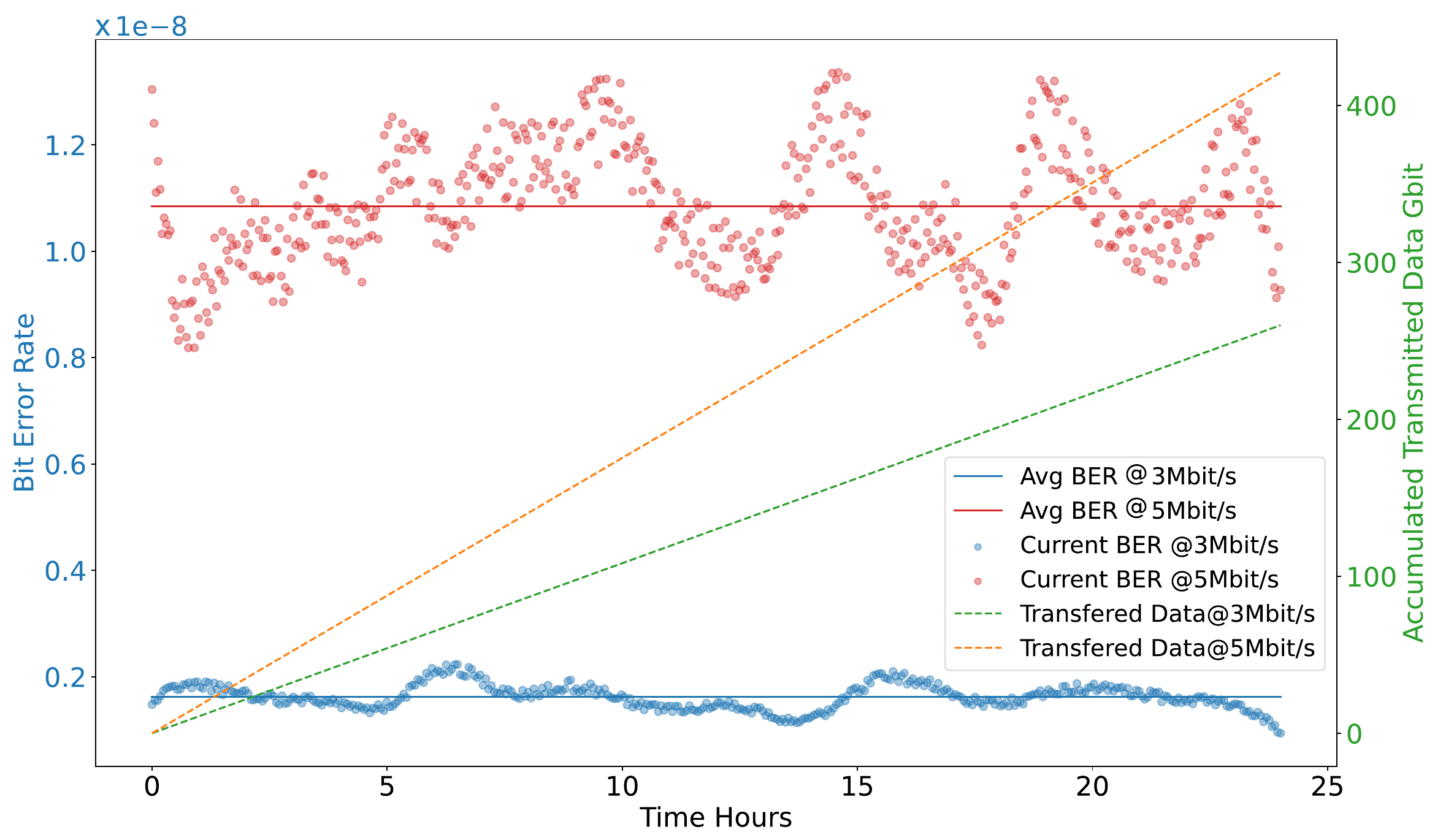}
    \caption{The diagram delineates the setup and outcomes of the Bit Error Rate (BER) testing procedure. The horizontal axis (X-axis) indicates the duration of testing, with each evaluation extending over 24 hours. The representation includes red dots and a corresponding red line, which illustrate the real-time BER measurements and the overall BER outcome, respectively, for a 5 Mbit/s transfer rate. Similarly, blue dots and a solid blue line depict the real-time BER observations and cumulative results for a 3 Mbit/s transfer rate. Additionally, the orange and green lines signify the total volume of data transmitted during the BER assessment, amounting to 260 Gbit for the 3 Mbit/s rate and 421 Gbit for the 5 Mbit/s rate. The visual also portrays the experimental arrangement, featuring the real-time decoded random image on the computer screen. Positioned at the left is the AD2 device serving as the receiver, while the right showcases the AD2 device functioning as the transmitter. Central to the setup is the tissue sample utilized for testing purposes.}
    \label{fig:BER_result}
\end{figure}

Figure~\ref{fig:BER_result} presents the results of the Bit Error Rate (BER) test conducted at transmission speeds of 5 Mbps and 3 Mbps, utilizing 10 mm bone and 7 mm skin as mediums. The figure illustrates the setup used for the BER assessment, including the real-time generation of a random image for testing purposes. Analysis of the data reveals that, at a transmission speed of 5 Mbps with Pulse Width Modulation (PWM), the BER was maintained below 1.09e$^-$$^8$. Conversely, employing a transmission speed of 3 Mbps with Pulse Density Modulation (PDM) resulted in a BER below 1.62e$^-$$^9$. The duration of the BER assessment exceeded 24 hours, during which more than 400 Gb of data were transmitted. The use of fresh tissue in these experiments imposed a limitation on the duration of the BER test due to the continuous dehydration of the tissue, which affects its optical and electrical properties. Observations from the test indicate that the BER exhibited fluctuations that appeared to follow a discernible pattern, potentially linked to variations in environmental lighting. However, due to the constrained timeframe of the experiment, a definitive correlation between environmental lighting conditions and BER fluctuations could not be established.
\section{Discussion}
The newly proposed wireless and leadless power and data system for endovascular brain-computer interfaces (eBCIs) offers a transformative solution in medical devices, particularly given the challenges associated with existing eBCIs. One of the principal advantages is the elimination of the long, cumbersome cable. By doing so, risks associated with physical connections, especially in vulnerable patient groups, are considerably reduced. Pediatric patients, for instance, who are in a state of continuous growth, would particularly benefit from a leadless system, as would patients with delicate blood vessels or conditions that render their vasculature susceptible to damage.

The enhanced data transfer capabilities of the proposed solution are another significant benefit. A transfer rate of over 2 Mbit/s represents a significant leap forward, allowing for the transmission of detailed electrocardiography (ECoG) data, which is invaluable for real-time monitoring and interventions. The robust power transfer system is another key advantage, with its capability to provide sufficient energy to the implant site while adhering to safety standards, ensuring both efficiency and patient safety.

However, several challenges arise when considering the implementation of this proposed solution. The task of miniaturizing the telemetry module to fit seamlessly within an endovascular stent introduces a series of challenges in electronic packaging. As medical devices become more compact, they often require more intricate designs and manufacturing techniques. This complexity can influence the reliability, longevity, and overall performance of the system, necessitating rigorous testing and iterations in the design phase.

Biocompatibility remains another major concern. While removing the long cable undoubtedly reduces certain risks, introducing novel electronic components and materials directly into critical regions like the bloodstream or brain tissue presents a new set of challenges. There's the potential for tissue reactions, unforeseen long-term effects, and other adverse events that can arise from the body's interaction with foreign materials. Ensuring that any newly introduced materials or design alterations are rigorously tested for biocompatibility is essential to circumvent potential complications and guarantee patient safety.

Moreover, while software simulations and modeling provide invaluable insights, they may not perfectly mirror real-world scenarios. Human physiology and tissue responses are complex, and the device's behavior under actual human conditions might present unanticipated challenges that simulations might not account for. Consequently, extensive in-vivo testing and post-implantation monitoring will be crucial to validate the system's performance and safety in real-world settings.

Finally, the leadless power and data solution for eBCIs offers a promising direction that addresses many of the drawbacks of current systems. However, it also introduces a new set of challenges that require careful consideration, rigorous testing, and continuous monitoring to ensure its viability, safety, and efficacy in practical applications.

\section{Conclusion}
In conclusion, this paper presents a novel optical wireless telemetry module designed for integration within a smart stent, aiming to overcome the limitations of current endovascular brain-computer interfaces (eBCIs) such as the Stentrode{\texttrademark}. By eliminating the need for a long wire connecting the stent electrodes to the encapsulated electronics, the proposed module enhances the eBCI's applicability for patients with weak blood vessels or susceptible vasculature and addresses the challenges associated with pediatric applications. 
The telemetry module, which can transfer data at maximum 5 Mbit/s and operate with less than 3 mW power consumption, demonstrates its potential through proof-of-concept experiments using discrete components and fresh bovine bone, muscle, and skin tissue. Future developments of application-specific integrated circuits (ASICs) will further refine the optical telemetry module's performance and pave the way for more versatile, safer, and less invasive eBCIs, ultimately transforming the landscape of neuroscience, engineering, and medical devices.

\section*{Acknowledgment}
\label{sec:ack}
The authors acknowledge the financial support from the Australian Research Council under Project DP230100019. The authors also acknowledge the support provided by the Sim4Life software support team.
\bibliography{Ref}
\end{document}